\documentclass[aps,pra,showpacs,twoside,twocolumn,10pt]{revtex4-1}
\usepackage[colorlinks=true, citecolor=blue, urlcolor=blue ]
{hyperref}
\usepackage{epsfig,newlfont,amssymb,amsfonts,amsmath,bm,subfigure,palatino,mathtools,amsthm,braket,soul,enumitem,color,graphics,graphicx,times,physics,bbold,comment, multirow, makecell,float}
\usepackage[normalem]{ulem}

\begin{document}

\title{
Discovering Factorization Surface of Quantum Spin Chains with Machine Learning 
}

\author{Nakul Aggarwal$^{1,2}$, Keshav Das Agarwal$^2$, Tanoy Kanti Konar$^2$, Leela Ganesh Chandra Lakkaraju$^2$, Aditi Sen(De)$^2$}

\affiliation{$^1$Department of Physics and Astronomy, Rutgers University, Piscataway, NJ 08854-8019 USA}
\affiliation{$^2$Harish Chandra Research Institute, A CI of Homi Bhabha National Institute,  Chhatnag Road, Jhunsi, Allahabad 211019, India}

\begin{abstract}
 Entanglement in quantum many-body systems is required for a variety of quantum information tasks, making it crucial to identify the parameter space in which the ground state is fully separable, known as the factorization surface (FS). Nonetheless,  the tuning parameters indicating FS for several quantum spin models remain unknown.  We employ symbolic regression (SR), a supervised learning technique, to determine a closed-form expression in the parameter regime corresponding to FS of quantum many-body Hamiltonians. We verify the effectiveness of this method by examining the analytically tractable models, namely a nearest-neighbor (NN) quantum transverse $XY$ model with additional Kaplan-Shekhtman-Entin-Aharony interactions, for which the FS is well-known. We construct an accurate expression for the FS of the $XYZ$ model by providing the parameter set through the SR algorithm in which the ground state  is derived by matrix product state formalism. 
 With a satisfactory level of accuracy, we estimate the FS for the long-range $XY$ model,  and the NN $XY$ model with Dzyaloshinskii–Moriya type asymmetric interaction for which the factorization surface is not known.

\end{abstract}

\maketitle

\section{Introduction}

Machine learning (ML) algorithms, both supervised and unsupervised, have been demonstrated  to be capable of forecasting counter-intuitive phases  in several quantum many-body systems for which order parameters are not easy to determine \cite{Carrasquilla_np_2017,Tanaka_jpsj_2017,vanNieuwenburg_np_2017,wang_prb_2017,Morningstar_jmlr_2018,cirac_2019_rmp}.  In  addition, these techniques can uncover the dynamical characteristics of the system by exploiting observables that can be measured in laboratories. Specifically,   recurrent neural networks capable of handling sequential data have been used to unearth the boundary of the many-body localization (MBL), and phase transition in random Heisenberg model \cite{van2018learning}, understand topological order by acting as a variational wavefunction ansatz \cite{hibat2023investigating, PhysRevResearch.2.023358}, reconstruct the ground state wavefunction of the \(XY\) model \cite{morawetz2021u} for analysis of various thermodynamic properties and iteratively, learn the ground state energy \cite{roth2020iterative} of large lattices. These results demonstrate ML as an appealing alternative  to tensor network methods like density matrix renormalization group (DMRG) \cite{schollwock2005density, schollwock2010}, and quantum Monte Carlo simulations \cite{qmc1, qmc2, qmc3}. Also neural network architecture is employed to detect entanglement in a many-body system \cite{bose_prl_2018}.



%


In contrast to deep learning, a supervised learning, known as symbolic regression (SR) \cite{Koza1994, Gerwin1974Sep, Langley1981Jan}, can be  deployed to acquire a compact analytical expression  for a given dataset, thereby revealing certain features of the system.  Despite the fact that SR has been shown to be NP-hard \cite{virgolin2022symbolic}, several methods have been proposed to tackle this problem ranging from genetic algorithms \cite{schmidt2009distilling} to deep learning \cite{udrescu2020ai}. Although it has never been utilized to investigate quantum spin models, it has been widely applied in astronomy to determine the mass scaling relations of black holes in spiral galaxies \cite{Davis2023DiscoveryOA}, to model galaxy-halo relations \cite{delgado2022modelling}, to rediscover Newton's law of gravitation in the solar system and determine the masses of celestial bodies  \cite{lemos2023rediscovering}, in mathematical physics to determine the precise relationships between topological invariants of hyperbolic knots \cite{craven2021disentangling} and predict Sasakian Hodge number $h^{2,1}$ \cite{aggarwal2024machine}, to learn bilateral accessibility in gravity models for international trade \cite{Verstyuk2022Mar}, to parameterize cloud cover in climate models \cite{grundner2023data}, and to find symbolic misfolding models for tau proteins in Alzheimer's disease \cite{ZHANG2024116647}.


On the other hand, the ground, thermal, and dynamical states of quantum spin models possess resources like quantum entanglement \cite{horodechki_rmp_2009} required for the development of quantum technologies like one-way quantum computer \cite{Raussendorf01, briegel_nature_2009}, quantum state transfer \cite{Bose03} and quantum metrology \cite{giovannetti_prl_2006}. More importantly, they can also be prepared as well as manipulated in laboratories using a variety of physical substrates such as trapped ions \cite{Murmann_prb_2015}, superconducting circuits \cite{WallraffPRX15}, photonic systems \cite{Pitsios_np_2017}, nuclear magnetic resonance \cite{Suter05, ChuangRMP05}, and cold atoms \cite{Bloch_np_2012}. However, it turns out that by manipulating the parameters of the model, one can reach  a surface, known as the factorization surface (FS) \cite{Kurmann1981Mar, Kurmann1982May, giampaolo_prl_2008}, in which the ground state is fully separable.  Nevertheless, obtaining this surface-equation proves to be challenging for both integrable and non-integrable models.  

To overcome this obstacle, we employ symbolic regression, in this work, to determine expressions of the factorization surfaces of multiple one-dimensional ($1D$) spin models with varying strata.
In particular,  we  focus on analytically solvable nearest-neighbor models, especially the transverse $XY$  spin models \cite{barouch_pra_1970} which may, additionally, involve either Dzyaloshinsky-Moriya (\(DUXY\)) \cite{dzyaloshinsky_1958} or Kaplan-Shekhtman-Entin-Aharony (\(KSEA\)) \cite{kaplan_jpb_1983,shekhtman_prl_1992} interaction. As non-integrable models, we examine  FS of the nearest-neighbor (NN) $XYZ$ spin chain in the presence of  an external magnetic field \cite{rossignoli_pra_2009_1} and long-range $XY$ models in which interaction strength obeys a power law  decay with the increase of coordination number \cite{cerezo_prb_2015,lakkaraju_pla_2021}.
It is noteworthy to mention that the factorization surfaces are known for the $XY$, \(KSEA\), and $XYZ$ models \cite{giampaolo_prl_2008, agarwal2023recognizing}; but it is unidentified for systems containing Dzyaloshinsky-Moriya  and long-range interactions.


By randomly sampling parameters of the Hamiltonian, we generate a dataset containing  entanglement between all possible neighboring spins.  
From this database, we collect all such parameters with  vanishingly small  entanglement content, which are then fed to the SR algorithm. In this work, we utilize the PySR package \cite{cranmer2023interpretable}  because, in contrast to other SR programs like AI Feynman \cite{udrescu2020ai}, it provides a number of benefits including custom expressions, parallelization, and full open-source availability. 
 By creating a tree of potential binary and unary operators, including the constants, this yields the closest equation. After pruning and adding additional branches from the permitted pool of operators and constants, it parses the tree and generates a few potential equations. The best possible equation can be achieved based on the performance measures, which are loss, score, and complexity, with the goal of maximizing score while minimizing loss and complexity. 

We exhibit that the factorization surfaces derived for the NN $XY$ and  \(KSEA\)  models via the PySR method and the analytical means do, in fact, match, thereby validating the efficacy of the machine learning approach. The parameter regime in which FS is known once more in the $XY$ model with Dzyaloshinskii-Moriya (\(DM\)) interaction agrees with the one derived from SR. Interestingly,  we  discover the FS equation in another parameter domain where FS is entirely unknown in the literature. 

By using the  data for the $XYZ$ Hamiltonian generated through the DMRG method with the matrix product state (MPS) ansatz,  we report that the output equation from the PySR algorithm replicates the well-known FS once more. When dealing with the long-range model, the data is produced using exact diagonalization (ED) for the appropriate parameter set. 
We observe that the PySR algorithm provides the expression of FS, known with other numerical and analytical methods for NN and next-nearest neighbor transverse $XY$ model,  within the acceptable levels of accuracy.


The  organization  of the paper is as follows: In Sec. \ref{sec:models}, we describe  various quantum spin models, including the ones that can be solved analytically and the ones that require numerical diagonalization. We also provide equations for their factorization surface, if they are known so that the effectiveness of SR in this problem can be confirmed. 
The comprehensive discussions on symbolic regression package, PySR (SubSec. \ref{sec:SR}),   steps involved in the implementation of the algorithm, and   the definitions of the  quantifiers assessing the performance of the algorithm are presented in Sec. \ref{sec:learning_fs}.  In Sec. \ref{sec:results}, we discuss the results obtained, starting from the integrable models to the non-integrable nearest-neighbor and long-range ones. We summarize in Sec. \ref{sec:conclusion}.



\section{Factorization surfaces of spin models: Analytically  vs Numerically solvable models}
\label{sec:models}

To achieve the goal of finding FS with the aid of machine learning techniques, we present here the concept of fully factorized states, and the spin models in which such states appear as ground states (GS)  by suitably tuning the parameters of the model.

\subsubsection{Fully factorized states}

A  $N$-party pure state is said to be  fully factorized  if it can be written as  \(\ket{\Psi_F}=\ket{\psi_1}\otimes\ket{\psi_2}\ldots \otimes\ket{\psi_N}\). To verify whether a given state is fully factorized, we compute its bipartite reduced density matrices after tracing out all the sites, except the first  and an arbitrary site $j$ (\(j=2, \ldots, N\)), given by $\rho_{1j} = \mbox{tr}_{\overline{1j}}(|\Psi_F\rangle\langle |\Psi_F|)$, where \(\overline{1j}\) denotes the sites except \(1\) and \(j\). We find the  entanglement  of $\rho_{1j}$   using logarithmic negativity \cite{horodecki_pla_1996, peres_prl_1996, vidal_pra_2002}, $\mathcal{E }=\sum_{j=2}^{N}\log_2{\left[2\mathcal{N}(\rho_{\text{1j}})+1\right]}$,
where $\mathcal{N}$ is the negativity of  $\rho_{\text{1j}}$, obtained by summing  over the absolute values of all the negative eigenvalues in the partially transposed state $\rho_{\text{1j}}^{T_j}$ \cite{peres_prl_1996, horodecki_pla_1996}. Note that in the case of nearest-neighbor  Hamiltonian,  nearest-neighbor sites, \(\rho_{12}\), typically possesses a significant amount of entanglement while entanglement in \(\rho_{1j}\) (\(j>2\)) is either negligibly small or vanishing and hence, we have \(\mathcal{E} = \log [2 \mathcal{N}(\rho_{12}) +1]\).  Towards finding FS, we check when $\mathcal{E}$ becomes vanishing small, i.e., when $\mathcal{E}<\epsilon $ (with \(\epsilon \sim \mathcal{O} (10^{-5})\)  being  a  number close to zero, decided from the accuracy of the computation performed). 

\subsubsection{Factorized states in analytically solvable spin models}

We consider a nearest-neighbor \(XY\) model with different symmetric and anti-symmetric exchange interactions, namely Dzyaloshinskii–Moriya   \cite{dzyaloshinsky_1958} and Kaplan-Shekhtman-Entin-Aharony  interaction \cite{kaplan_jpb_1983,shekhtman_prl_1992} in the presence of  an transverse magnetic field. In  these models, eigenenergies and eigenstates can  be obtained analytically for an arbitrary number of sites and also in the thermodynamic limit (\(N \rightarrow \infty\)). 

\textit{Transverse \(XY\) spin chain.} For a system of $N$ interacting spin-$1/2$ particles on a one-dimensional lattice, the \(XY\) Hamiltonian is written as
\begin{equation}
\begin{split}
    H_{XY}=\frac{1}{2}\sum\limits_{j=1}^{N}&\Big[J\left\{\frac{1+\gamma}{2}\sigma^x_{j}\sigma^x_{j+1}+\frac{1-\gamma}{2}\sigma^y_j\sigma^y_{j+1}\right\}+h'\sigma^z_j\Big ],
\end{split}
\end{equation}
where $\gamma \in [0,1]$ is the  anisotropy parameter, $J<0$ and $h'$ represent the strengths of ferromagnetic nearest-neighbor interaction in the \(xy\)-plane and the external magnetic field respectively, and $\sigma^\alpha$ ($\alpha=\{x,y,z\}$) are the Pauli matrices. Here we consider periodic boundary conditions (PBC) such that $\sigma^\alpha_{N+1}=\sigma^\alpha_{1}$ and set $h'/J=h$. Using Jordan Wigner and Fourier transformation, the Hamiltonian $H_{XY}$ can be mapped to a spinless fermionic model (see Appendix \ref{app:xy_spin_model}). This model can be simulated in different trapped ions and cold-atom platforms \cite{cirac_iontrap_XY}. It can be analytically shown that the doubly degenerate ground state of this model becomes fully factorized (see Appendix \ref{app:fact_surface_calc}) when the parameters satisfy the surface equation, given by $ h^2+\gamma^2=1$, which is known as the factorization surface \cite{giampaolo_prl_2008}.
There is a  prescription to find the FS of a given model using the single qubit unitary method \cite{rossignoli_pra_2008, giampaolo_prl_2008, giampaolo_prb_2009, rossignoli_pra_2009}.

\textit{\(XY\)  spin chain with \(KSEA\) interactions.} Let us now add \(KSEA\) interaction in the transverse \(XY\) spin chain \cite{kaplan_jpb_1983},
\begin{equation}
    H_{KSEA}= H_{XY} + \sum\limits_{i=1}^{N}\frac{k'}{4}\left(\sigma^x_{i}\sigma^y_{i+1}+\sigma^y_{i}\sigma^x_{i+1}\right),
\end{equation}
where $k = k'/J$ is the strength of the symmetrical helical \(KSEA\) interaction \cite{kaplan_jpb_1983,shekhtman_prl_1992,zheludev_prl_1998}. It has been found that it exists in a helimagnet like $Ba_2CuGe_2O_7$ \cite{chovan2013field} and can successfully explain the weak ferromagnetic behavior of $La_2CuO_4$ \cite{tsukada2003significant}. In this case,  the equation for the FS reads $h^2+k^2+\gamma^2=1$ which can also be analytically proven (see Appendix \ref{app:fact_surface_calc}).


\textit{\(XY\)  spin chain with \(DM\) interactions (\(DUXY\) model).}
The \(XY\) Hamiltonian with Dzyaloshinskii and Moriya interaction can be represented as
\begin{equation}
H_{DUXY}=H_{XY}+\sum\limits_{i=1}^{N}\frac{d'}{4}\left(\sigma^x_{i}\sigma^y_{i+1}-\sigma^y_{i}\sigma^x_{i+1}\right),
\end{equation}
where $d= d'/J$ is the strength of the anti-symmetric exchange interaction. Such a term successfully explains the presence of weak ferromagnetism in certain types of materials like \(\alpha\)-\(\text{Fe}_2\text{O}_3\), \(\text{Mn}_2\text{CO}_3\) which breaks the mirror symmetry in the models \cite{dzyaloshinsky_1958,  PhysRevLett.4.228, PhysRev.120.91, PhysRev.115.2, dmitrienko_nature_2014,pattanayak_prb_2017}. 
In the \(DUXY\) model, for $d<\gamma$, the factorization surface is independent of the strength of  $d$ and is identical to that of the \(XY\) model. However, for $d>\gamma$, there does not exist a closed and compact expression representing the factorization surface. The effectiveness of the ML method will first be tested with the models for which FS equations are known and then we apply it to the \(DUXY\) chain with $d>\gamma$.

\begin{figure*}
    \centering    \includegraphics[width=\textwidth]{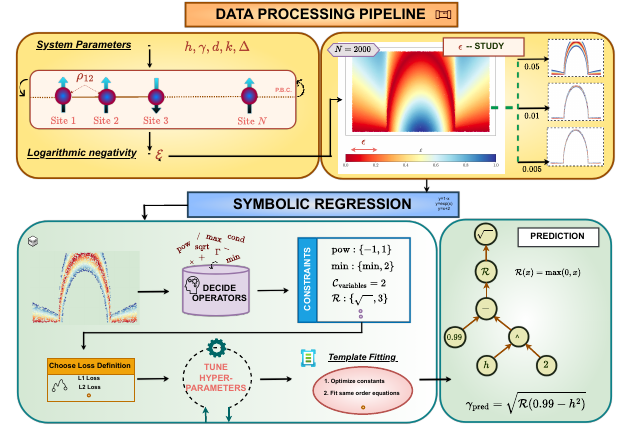}
    \caption{\textbf{Data processing pipeline.} We first  compute $\mathcal{E}$  of the spin chain model under consideration by varying the system parameters. For the \(XY\) and \(KSEA\) spin models, we perform the $\epsilon$-study, critical of the precision calculation of $\mathcal{E}$. For the $XYZ$ model,  we create a random dataset by varying the system parameters which is then fed to the SR. \textbf{SR Pipeline.} We select the operators based on domain knowledge of the spin models,  operator constraints, and the loss function, and by tuning the relevant hyperparameters, we predict equations. In case there is a further requirement  to improve the equation, we implement template fitting to improve upon the prediction. As we will show, the last strategy is well-suited for the $DUXY$ spin model, where we do not know the FS equation.}
    \label{fig:enter-label}
\end{figure*}

\subsection{Finite-sized spin chain}

\textit{$XYZ$ spin model.} The $XYZ$ model with open boundary conditions hosts unequal strengths of interaction in all three directions, given by
\begin{equation}
    H_{XYZ}= H_{XY}+\frac{\Delta'}{4}\sum_i^{N-1}\sigma^z_i\sigma^z_{i+1},
    \label{eq:XYZ_ham}
\end{equation}
where $\Delta = \Delta'/J$ is the coupling constant in the $z$-direction. This model has been shown as an effective description of various many-body systems \cite{rmp_vedral, QPT_book_2015}, and can be simulated in various quantum computing hardwares like quantum dots \cite{divenco_pra_1998}, optical lattices \cite{sen_ad_phys_2007}, trapped ions \cite{blatt_phys_rep_2007, sug_bose_pra_2008} and superconducting circuits \cite{superc_nature}. For $\gamma = 0$ and $ \Delta = J$, the model reduces to the Heisenberg model which can be analytically solved using Bethe ansatz although for $\gamma \ne 0$ and $ \Delta \ne J$, the ground state can  be found by numerical methods \cite{schollwock2005density, schollwock2010}  which follow the area law of entanglement \cite{schollwock2010}. In this work, we use the ITensor package \cite{itensor} to obtain the GS for various parameters, selected randomly from $\gamma\in [0,1]$, $\Delta\in [0,1]$ and $h\in [0,2.5]$.
To minimize the boundary effect, we compute entanglement between two sites, namely, between $\frac{N}{2}$ and $\frac{N}{2}+j$ spins $(j = 1,2, \ldots, \frac{N}{2})$. Although the GS for arbitrary sites of this model cannot be obtained, the FS \cite{rossignoli_pra_2009}  can be shown to follow the equation, given by $h^2+\gamma^2={(1+\Delta)^2}$. 
    
\textit{Long-range $XY$ spin chain.}
Long-range Hamiltonian occurs naturally in many experimental setups such as Rydberg atom arrays, dipolar systems, trapped-ion setups, and cold atoms in cavities \cite{Lahaye_2009, saffman_rmp_atoms, Jurcevic2014, monroe_sa1, monroe_sa2, defnu_rmp_review}. These long-range interactions results in non-local behavior in the ground state properties \cite{Vodola_2016, wang_prx_lr} and in the topological as well as dynamical features \cite{cirac_prl_lr}. The long-range $XY$ Hamiltonian $H_{LR}$ can be written  as
\begin{eqnarray}
    \nonumber H_{LR}=\frac{1}{2}\sum\limits_{i=1}^{N}\sum\limits_{j=1}^{\mathcal{Z}}&\Big [J^\prime\left\{\frac{1+\gamma}{2}\sigma^x_{i}\sigma^x_{i+j}+\frac{1-\gamma}{2}\sigma^y_i\sigma^y_{i+j}\right\}\\
    & + h^\prime \sigma^z_i\Big ], 
\end{eqnarray}
where $J^\prime = -\frac{J}{j^\alpha}$ indicates the ferromagnetic interaction strength,  $\mathcal{Z}$ is the coordination number and the interaction strength between spins follows a power-law decay, parameterized by $\alpha$. We again set \(h'/J = h\). 
The expression for the factorized surface of this model reads as $h=\sqrt{1-\gamma^2}\sum_{j=1}^\mathcal{Z} \frac{1}{j^\alpha}$. 

\section{Symbolic Regression and its implementation details}
\label{sec:learning_fs}

The machine learning technique that we  deploy here to obtain the FS is the symbolic regression, which we discuss along with the justification for this choice over traditional algorithms.  We then present the implementation of SR step by step. 

\subsection{Symbolic Regression}
\label{sec:SR}
For the reliable estimation of an empirical relationship between the input features and the output of a high-dimensional scientific dataset, it is pertinent to look for interpretable equations that can conform to the symmetries and constraints inherent in the system being studied. At the same time, SR should be highly capable in handling three important criteria -- (i) detecting discontinuities in the equations, (ii) stand robust to any potential noise or outliers and (iii) remain valid across various limits of the dependent variables. To illustrate this idea, consider the FS equation of \(KSEA\), given by $\gamma=\sqrt{1-h^2-k^2}$. The argument inside square root must be positive and hence SR must output an equation of the form $\gamma=\sqrt{\max(0,1-h^2-k^2)}$, thereby tackling discontinuities by incorporating the maximum operator ``\(\max\)". In the limit $k\to 0$, we should also recover the  FS of the \(XY\) model, which can keep it 
a outlier-safe. These outliers can exist due to numerical errors during the calculation of $\mathcal{E}$.  

Having setup the required SR's norms above, we are now in a capacity to briefly explain SR's working principles. In SR, the equations are constructed from operators, input features, and constants. Once the operator pool and their respective constraints are specified using domain knowledge, SR can recover existing equations and discover novel ones to the highest degree of accuracy possible. Even in cases where certain features of the system are not completely known, it is possible to predict multiple equations of different complexities (which we will define later) using generic operators that fit the data reasonably well. This is then followed by selecting equations that provide the deepest insight into the system and are simple in terms of complexity, in alignment with Occam's razor principle. There exist many algorithms and software to perform SR from sparse regression to genetic modeling. In particular, PySR \cite{cranmer2023interpretable} is an open-source Python package based on genetic programming and dedicated specifically to scientific symbolic regression.  
 Another popular SR software tool named AI Feynman \cite{udrescu2020ai} fuses elements of dimensional analysis, polynomial fit, brute-force methods, and deep learning inspired techniques to distill equations. In comparison to Eureqa \cite{schmidt2009distilling}, another SR package, although commercial, AI Feynman discovered $100\%$ of the Feynman equations. To select the most suitable SR candidate for this study, we require certain additional conditions like the versatility to define custom loss functions,  discontinuous and custom operators. PySR meets these particular criteria very well \cite{cranmer2023interpretable} which AI Feynman does not. 
In PySR, expressions are represented as trees. The number of binary trees with $n$ vertices is $(2n)!/((n+1)!n!)$. Since the underlying equation for the phase space is very large, it is prominently suitable for smaller low-dimensional datasets, in particular, for a dataset size of $\lesssim 10,000$ training points and dimensions of the input feature space should be ideally less than $10$ \cite{wadekar2023augmenting}. In this work, the feature space will be at most three-dimensional. PySR outputs equations at various complexity levels upto a maximum value which is defined by the user. The equation complexity, \(\mathcal{C}\), is determined by the total count of operators, variables and constants in the equation. Since, this is a hyperparameter, one has to specify its maximum value. The maximum user-defined complexity is called `maxsize'. Maxsize has to be reasonably selected in accordance with the number of input features and some prior understanding of the system. Setting maxsize too high can easily lead to an overfitted complex equation which can be hard to interpret. Conversely, if it is too small in value, the predicted equation might not represent the system accurately. It also allows to set constraints on nesting of operators like $\mathrm{max}:\{\mathrm{max}: 2, \mathrm{min}: 1\}$ -- it  indicates that inside the \(\max\) operator, there can be atmost two \(\max\) operators and one minimum operator ``\(\min\)". Nesting helps to control overfitting, recursion of operators multiple times and can guide PySR to the right equation sub-space.    

Let us now justify the reasons behind choosing SR instead of traditional curve fitting or deep neural network. 
In case of curve fitting (CF), it assumes a specific functional form of the output parametrized by a set of variables. While in SR, such a prior requirement is avoidable since the underlying input space is a collection of possible equations including the possibility of optimizing the constants. In the scenario of deep neural networks, it can learn an accurate representation of the output as a function of the important features, although an explicit form cannot be extracted. Further, they are not generalizable and typically hard to interpret.

 
\subsection{Data generation and types of studies}

Before presenting the results, let us first describe in details how we generate the data sets for entanglement of different spin models, which are then fed to SR. We perform two kinds of studies -- (1) $\epsilon$-catalogue --  we collect nearest-neighbor $\mathcal{E}$ values of the ground state for the \(XY\) and the \(KSEA\) models when it is less than the fixed value of, say $\epsilon \sim 10^{-4}$ and these values are then inserted into the ML algorithm. Here, $\epsilon$ is a measure of numerical precision during the evaluation of $\mathcal{E}$. (2) Secondly, we analyze how algorithm's performance varies with the size of the data set. The quantifiers used to determine whether the equation describing specific system properties is the best one are further explained.

To investigate the FS of the given model, specifically we perform the $\epsilon$-catalogue by adopting the following procedure:
\begin{itemize}
    \item \textit{Step 1}. We randomly generate $30$ million points from the parameter space of a given model. For example, the sets $\{h_i, \gamma_i\}$ for the \(XY\) model and additional $\{k_i\}$ and $\{d_i\}$ values for the $H_{KSEA}$ and $H_{DUXY}$ models respectively are sampled randomly, where the cardinality of the set parametrized by $i$ is of the order of millions for models that can be solved analytically and in thousands for the spin models that can only be solved by numerical methods, for example, the \(XYZ\) model and the long-range \(XY\) model. In all these models, the canonical equilibrium state, $\rho=\frac{e^{-\beta H}}{Z}$, with $Z = \text{tr}(e^{-\beta H})$ (partition function), $\beta=\frac{1}{k_B T}$, $k_B$ being the Boltzmann constant having temperature $T$ is constructed with $\beta > 100$ and the corresponding nearest-neighbor entanglement, $\mathcal{E}$, is computed after tracing out $N-2$ sites for each set of parameters $\{h_i, \gamma_i, k_i, d_i\}$
    with a total of $N=2000$ sites. We notice that $\mathcal{E}$ obtained with $N=2000$  remains intact with the increase of \(N\) and hence it can be considered as the  thermodynamic limit. 

    \item \textit{Step 2}.  We use the entanglement dateset to identify a $\epsilon$-radius zone where the FS equation can be predicted. We are interested in finding out how far we can dependably extract the correct equation for the known FS, as shown by the parameter $\epsilon$, which should be taken up to be the numerical precision. We extract $\epsilon$, from the following collection, $\{0.0001, 0.005, 0.001, 0.005, 0.01, 0.05\}$.
    
    \item \textit{Step 3}. For each $\epsilon$, we initially select a random sub-sample of $3000$ data points and tune SR algorithm's hyperparameters using the package ``Hyperopt" \cite{bergstra2015hyperopt}. This is crucial in identifying several relevant hyperparameters like the right batch size (the dataset is divided into smaller random sub-datasets of size ``batch-size"), maxsize, constraints and the operator pool.  This is done to reduce overfitting and decrease training time.

    \item \textit{Step 4}. Subsequently, again for a given $\epsilon$, we create $30$ random training datasets of $2000$ and $4000$ points for every $\epsilon$ for the \(XY\) and the \(KSEA\) models respectively. Since \(KSEA\) has one additional parameter, we double the dataset size to account for thorough exploration of the parameter space. PySR is then trained on this sample to obtain the predicted equation with the hyperparameter values as evaluated in \textit{Step 3}. 
    \item \textit{Step 5}. For every such dataset, equations of FS are predicted and their respective loss, score, and complexity values  are calculated for assessing the performance of the SR algorithm. In order to aggregate these results, we report their respective averages.
\end{itemize}

In order to train the model, we define the  weighted mean-square error loss, given as 
\begin{equation}
\mathcal{L}=\sum\limits_{i=1}^{N_{\text{train}}}w_i\left(\gamma^{\text{PySR}}_i-\gamma^{\text{truth}}_i\right)^2,
\label{eq:loss}
\end{equation}
where $\gamma^{\text{PySR}}_i$ and $\gamma^{\text{truth}}_i$ are the PySR's prediction and the actual $\gamma$ values for the $\mathrm{i}^{\mathrm{th}}$ training point respectively, $N_{\text{train}}$ is the number of samples in the training dataset and $w_i$ is the weight of each point. To quantify the usefulness of a predicted equation, PySR defines a metric called `Score',  $\mathcal{S}$, defined by  
\begin{equation}
    \mathcal{S}=-\frac{\delta \ln{\mathcal{L}}}{\delta \mathcal{C}}.
    \label{eq:score}
\end{equation}
The score $\mathcal{S}$ penalizes equations at a higher complexity while trying to minimizing the loss $\mathcal{L}$. The purpose of $\mathcal{S}$ is to punish highly complicated equations. We restrict the maximum complexity $\mathcal{C}$ in proportion to the number of input features. There is thus a trade-off between score and loss. Although there can exist an equation with a high score but it can compromise loss. To take this into account, PySR has an inbuilt model selection criterion, called ``Best" which outputs an equation having the highest score in the candidate pool of the equations within $1.5$ times the least loss equation. We show the results of both the ``Highest Score" (for brevity, written as ``HS") and ``Best" equations for each $\epsilon$. 

For both the models, we select the following pool of unary operators, $\{\mathrm{square}, \mathrm{cube}, \mathrm{sqrt}\}$ and binary operators $\{\mathrm{max}, \mathrm{min}, +, -, *\}$. As the name suggests, the unary and binary operators have one and two arguments respectively. The maximum and minimum discontinuous operators are chosen to account for any plateaus or discontinuous sub-regions in FS. We also perform general but not too restrictive nesting on these operators. For example, the maxsize for the \(XY\) model is set to $15$ while  for the \(KSEA\) spin chain,  maxsize of $25$ is taken as there are  two input variables, $h$ and $k$, instead of a single one in the \(XY\) model. Naturally, we suspect the complexity to increase for the  \(KSEA\) model. Since the datasets for both the \(XY\) and the \(KSEA\) models  have roughly similar parameter space-density, each point has a similar number of points in its neighborhood, barring some points at the periphery of the parameter space. This allows us to set the weights in the loss function to be unity. The batch-size for both the spin models is optimized to $64$.

 For the \(XYZ\) model, we perform a similar analysis although we change the number of sampled states in \textit{Step 3} by keeping the $\epsilon$ constant in \textit{Step 2} to $0.01$. Here, we initially observe that $N_{\mathrm{train}}=2000$ is not enough to find a reliable equation which fits the \(XYZ\) dataset well. Thus, it becomes important to study the \(XYZ\) model from the perspective of changing size of the training dataset. We predict equations for the three cases: $N_{\mathrm{train}}=2000, 4000$ and $6000$. The operator pool and maxsize are identical to that of the \(KSEA\) model.  
 
 In case of the \(DUXY\) spin chain, there are two important partitions of the dataset: $d<\gamma$ where the FS is well-established and its counterpart $d>\gamma$ for which the FS equation is unknown. For the former case, we keep the parameters same as in the \(KSEA\) model, given the same number of input features and train PySR to predict its equation. After showing proof of concept in the  case of \(H_{XY}\), \(H_{KSEA}\) and \(H_{XYZ}\), the case of \(DUXY\) with $d>\gamma$ serves as a perfect testbed to investigate the capability of PySR. For this model, $N_{\text{train}}=10^4$, $\epsilon=5\cdot 10^{-4}$ and batch-size is optimized to $128$. We apply following two different techniques to output a FS equation for the \(DUXY\) model in the domain of $d>\gamma$:
 \begin{itemize}
     \item \textit{PySR's preliminary prediction.} We predict the expression of the FS with the same PySR's settings as that of the \(KSEA\) spin chain. 
     \item \textit{PySR + Template fitting.} We now implement a simple curve fitting technique  based on the functional form of the default equation towards optimizing the predicted equation by PySR. The method is known as template fitting (discussed in the next subsection).  
 \end{itemize}
 For each technique, we check the prediction across various slices along the set of parameters, $d=\{0,1.0,2.0\}$. This developed methodology may help to scan a gamut of potential equations. 
\begin{figure*}
    \centering
    \includegraphics[width=\textwidth]{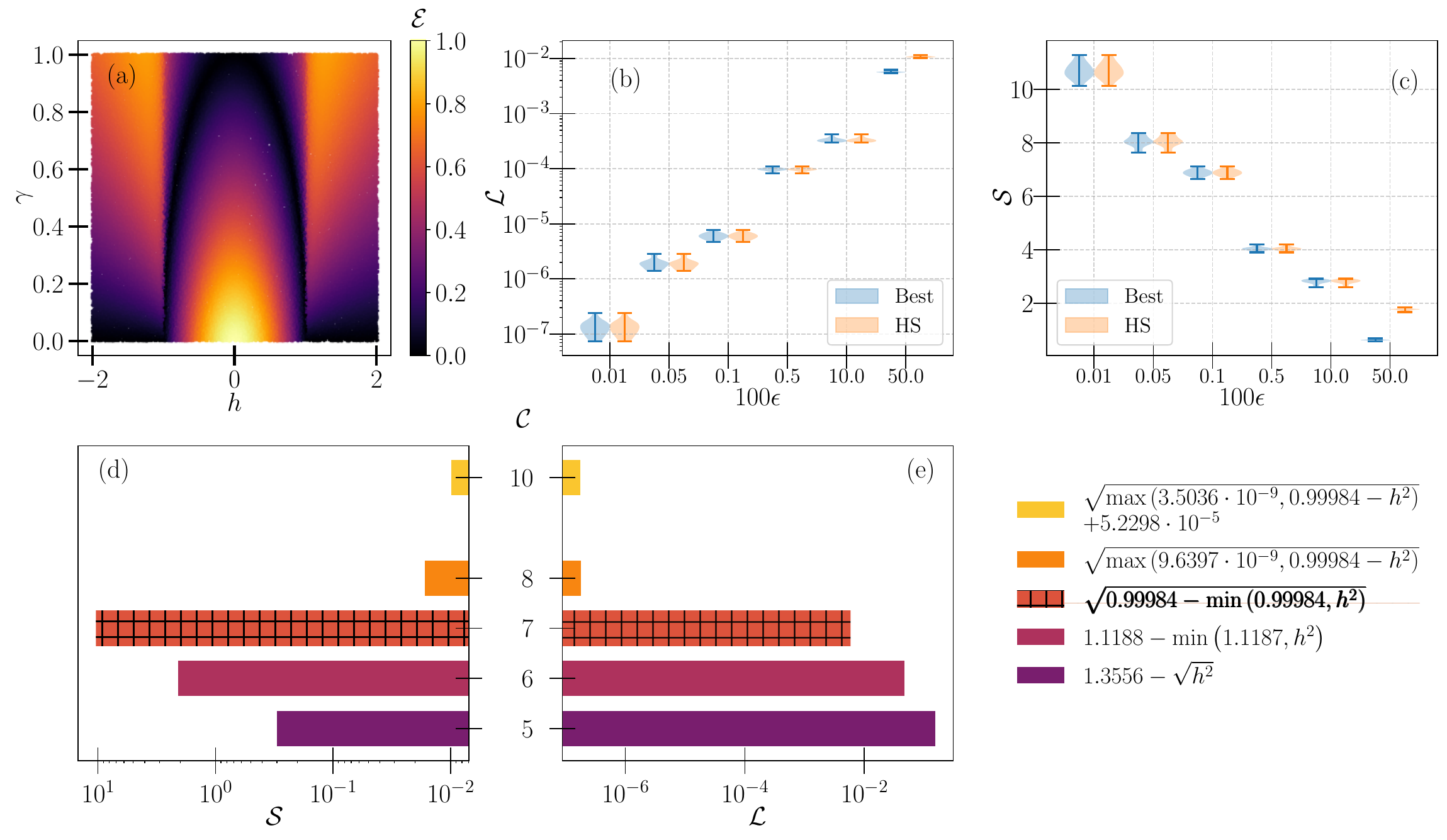}
    \caption{\textbf{Symbolic regression results for the \(XY\) model.} (a) Bipartite entanglement, $\mathcal{E}$ as a function of parameters, $h$ (abscissa) and $\gamma$ (ordinate). The FS can be clearly seen as a black curve whose equation is $\gamma=\sqrt{1-h^2}$ for $-1\le h \le 1$. Here $N=2000$. (b)-(c) Loss $\mathcal{L}$ and score $\mathcal{S}$ against $100\epsilon$ respectively.  It is evident that the Best and HS equations have similar functional forms at each $\epsilon$. The loss is significantly higher at $\epsilon\ge 0.01$ where the dataset is contaminated by points not representing FS.  (d)-(e) Optimized equations at each complexity level $\mathcal{C}$ for $\epsilon=10^{-4}$. Both Best and HS equations are identical here and  displayed in bold: $\gamma=\sqrt{0.99984-\mathrm{min}(0.99984,h^2)}$. This emphasizes that SR algorithm is successful in discovering the right FS equation for the \(XY\) model. All the axes are dimensionless.  }
    \label{fig:XY_9_plot}
\end{figure*}

\textit{Long-range \(XY\) spin model}:  In the long-range \(XY\)  spin model, \(H_{LR}\), we want to predict an equation for $\gamma$ as a function of $\alpha$ and $h$. The long-range FS can be written as
\begin{equation}
    \gamma = \sqrt{\left(1-\frac{h^2}{R_{\text{sum}}^2}\right)},
\end{equation}
where $R_{\text{sum}}$ is given by
 \(   \mathcal{R}_{\text{sum}}
=\sum\limits_{j=1}^{\mathcal{Z}}\underbrace{\frac{1}{j^\alpha}}_{\mathcal{R}_j}\) \cite{defnu_rmp_review}.
Predicting an equation for long-range \(XY\) spin model is challenging as each subsequent term in $R_{\text{sum}}$ is more suppressed than the previous one for $\alpha \geq 1$. Additionally, 
there can be multiple equations which could fit the dataset very well, thereby possibly representing the FS. 
However, some of them can be discarded with the help of the knowledge of the system. 
For example, 
let us assume that we  apriori  know that the model has spin-spin interactions following a power-law decay (i.e., which is a function of \(\alpha\)).   This allows us to identify the FS  equation as a function of $\alpha$ through the terms $\mathcal{R}_j$ which represents the range of interactions present in the Hamiltonian, denoted by \(\mathcal{Z}\).

\subsection{Template Fitting}

After PySR predicts an equation, this equation can be optimized further using a simple curve fitting. In case of determining FS, we first investigate whether the plot of true $\gamma$ vs predicted $\gamma$ follows a linear trend along the line $y=x$. If there are deviations from this line, we can improve upon the equation from SR by first polishing the values of the constants in the equation. Subsequently, we can try fitting an equation of the same functional form as that output by SR. 

Let us illustrate this with an example. Consider the true relationship for $z$-dependent on two variables $x_1$, and $x_2$ such that $z(x_1,x_2)=10x_1^{2.5}x_2^{3.2}$. Suppose PySR outputs $z(x_1,x_2)=2.1x_1^{1.2}x_2^{2.3}$ and we find that this equation does not fit the dataset well. Since PySR predicts a power law form, we can now fit an equation of the same functional form as $Ax_1^{B}x_2^{C}$ via curve-fitting and obtain the constants $A$, $B$ and $C$. We follow this strategy to predict a potential FS equation of the \(DUXY\) model for $d>\gamma$ whose equation does not yet exist in the literature.

\section{Estimating Factorization surface in nearest-neighbor models through SR}
\label{sec:results}

In order to construct factorization surface of the NN spin models, we present the scenario in two ways  -- (1) let us consider the models for which FS is known \cite{rossignoli_pra_2008} so that we can confirm the effectiveness of the method used. Among all these models, we first carry out the analysis to obtain the FS of the \(XY\) model for different values of system parameters $\{h_i, \gamma_i\}$. By increasing the number of system parameters from two to three, $\{h_i, \gamma_i, k_i\} $ and $\{h_i, \gamma_i, \Delta_i\} $, we reach the FS equation of the \(KSEA\) and \(XYZ\) models respectively. (2) We then apply our method to the \(DUXY\) model with the set of parameters $\{h_i, \gamma_i, d_i\} $ for which analytical expression for the FS is not known in the literature \cite{roy_prb_2019}. 

\subsection{FS of the transverse \(XY\) and \(KSEA\) model reached via SR}

By analyzing the pattern of  nearest-neighbor entanglement of the transverse \(XY\) model in  $(h,\gamma)$-plane, it is apparent that for $\mathcal{E} < 10^{-4}$, the square of the strength of the magnetic field is related to the anisotropy parameter via a polynomial equation of degree $2$ (see Fig. \ref{fig:XY_9_plot}(a)). Further, at $|h|=1$, the FS becomes discontinuous. Both the features indicate that the vanishing entanglement occurs on the half-circle in the $(h,\gamma)$-plane. By using different measures, let us now establish that the nearest-neighbor entanglement values obtained via PySR by varying $\gamma$ and $h$ are appropriate. 

{\it{True vs predicted values}}: When $\gamma^{\text{true}}$ and $\gamma^{\text{PySR}}$ coincide, i.e., $\gamma^{\text{true}} \equiv \gamma^{\text{PySR}}$, the algorithm can correctly infer the equation for the FS. We show in Fig. \ref{fig:XY_9_plot}(b) that this is indeed the case with $\epsilon < 10^{-4}$ where $\mathcal{L}<10^{-6}$. We find that except at the quantum critical point, i.e., at $|h|=1$, the error between $\gamma^{\text{true}}$ and $\gamma^{\text{PySR}}$ is negligible for all values of $h$. The error becomes higher at $|h|=1$ due to the sudden discontinuity in the FS. Our observations endorse that PySR  works satisfactorily and the only difference arises from the values of the constants in the FS equation.
\begin{figure*}
    \centering
    \includegraphics[width=\textwidth]{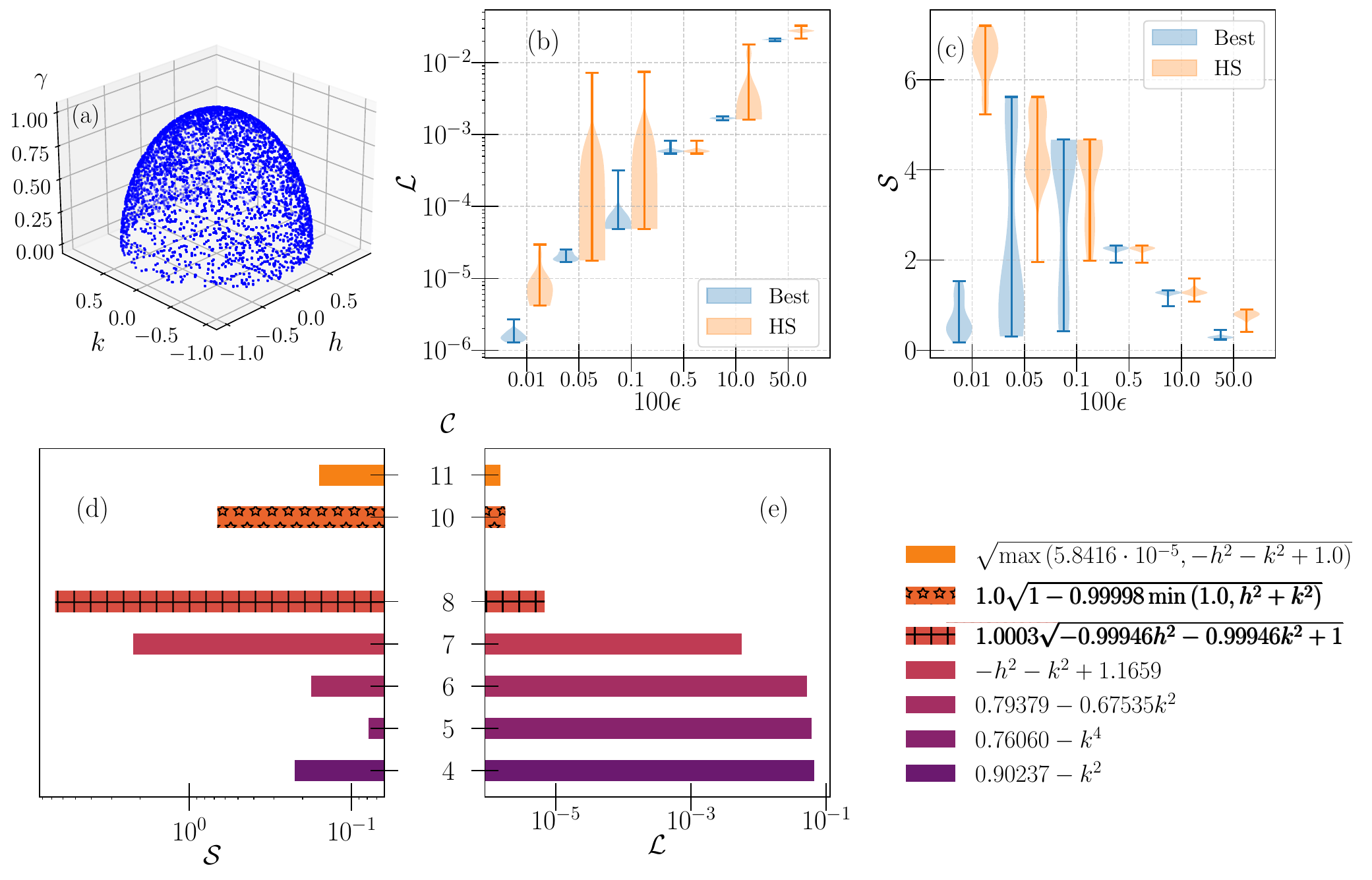}
    \caption{\textbf{Symbolic regression results for the \(KSEA\) model}. (a) The surface in terms of $h$ (\(x\)-axis), $k$ (\(y\)-axis) and $\gamma$ (\(z\)-axis) when bipartite entanglement $\mathcal{E} < 10^{-4}$   in \(H_{KSEA}\). (b)-(c) Loss $\mathcal{L}$ and score $\mathcal{S}$ vs. $100\epsilon$ respectively. The Best and HS equations resemble to some degree only at $\epsilon=10^{-4}$. However, the HS equation deviates quite dramatically from the Best one at increasing values of $\epsilon$. (d)-(e) Optimized equations at each complexity level $\mathcal{C}$ for $\epsilon=10^{-4}$. Both Best and HS equations are different here and displayed in bold. The Best equation is $\sqrt{1-0.99998\text{min}(1,h^2+k^2)}$. This again reiterates that SR algorithm is a lucrative method to establish FS equations. All the axes are dimensionless.  }
    \label{fig:KSEA_9_plot}
\end{figure*}

{\it Certifying FS via loss, score and complexity.} By exploiting different metrics, $\mathcal{L}$, $\mathcal{S}$ and $\mathcal{C}$ defined in Eqs. (\ref{eq:loss}), and (\ref{eq:score}), we now check whether the predicted values from PySR and the true values of $\gamma$ are certainly close to each other for several values of $\epsilon$-set in the algorithm. As $\epsilon$ increases, we move further away from FS, and the loss of the predicted equation also grows. Till $\epsilon=10^{-3}$, $\mathcal{L}$ is below $10^{-5}$ which is acceptable, while for $\epsilon\ge 0.005$, $\mathcal{L}$ increases. Furthermore, we note that although the variance across several rounds remains small, all random datasets perform poorly at this $\epsilon$. The story of $\mathcal{S}$ is similar to that of $\mathcal{L}$ but in reverse (see Fig. \ref{fig:XY_9_plot}(c)). For e.g., $\mathcal{S}$ is greater than $10$ when $\epsilon=10^{-4}$. There is a striking difference at $\epsilon=0.05$, where the interquartile range for the HS equation increases. This suggests that PySR can find equations with a high overall score but the respective loss is simultaneously higher as well. It highlights the importance of carrying out the comparison between $\mathcal{L}$ and $\mathcal{S}$ in juxtaposition and justifies the usage of the Best equation metric. We also try to see variations across  the complexity, $\mathcal{C}$, in equations against different $\epsilon$, but instead find its value as $6$ in each round on average for each $\epsilon$. This underscores the fact that FS remains interpretable at all orders in $\epsilon$. 

For a fixed $\epsilon = 10^{-4}$ value, let us find the best equations obtained for characterizing the FS equation. Corresponding to each equation, we find complexity, $\mathcal{C}$ and its score, $\mathcal{S}$ and loss, \(\mathcal{L}\) as illustrated in Figs. \ref{fig:XY_9_plot}(d)-(e). Comparing these two figures, we again understand the complementary behavior of $\mathcal{S}$ and $\mathcal{L}$. Since in this scenario, the FS is known analytically, we clearly find that the best equation describing the FS is given by \begin{equation}
\gamma=\sqrt{0.99984-\mathrm{min}(0.99984,h^2)}.
    \label{eq:XY_FS_PySR}
\end{equation} 
We notice that it is also the highest score equation. We can observe that PySR remarkably predicts equations for FS at a loss lower than $10^{-6}$, although this occurs at the added expense of higher equation complexity. We want to avoid the very complicated looking equations as they are usually a by-product of overfitting. Although Eq. (\ref{eq:XY_FS_PySR}) has higher loss than some other equations, but it is still a pretty decent loss at a lower complexity when compared to its counterparts. 

\renewcommand{\arraystretch}{1.4}
\begin{table*}[]
    \centering
    \begin{tabular}{c@{\hspace{2em}} c c @{\hspace{2em}} c @{\hspace{2em}} c@{\hspace{2em}}c }
    \Xhline{2\arrayrulewidth}
    \Xhline{2\arrayrulewidth}
    Training Size & Model Criterion & Equation & Complexity & Loss & Score \\
    \Xhline{2\arrayrulewidth}
    \Xhline{2\arrayrulewidth}
    \hspace{2 em}\multirow{2}{4em}{$2000$} & Best & $\sqrt{\mathrm{max}\left(0.0119,1.2145-(-\Delta+h)^2\right)}$ &	8 & $0.0134$ & $0.6371$ \\
    & HS & $1.5605 - \sqrt{h}$ & 4 & $0.0589$ & $0.7900$ \\
    \Xhline{2\arrayrulewidth}
    \hspace{2 em}\multirow{2}{4em}{$4000$}  & Best & $\sqrt{\mathrm{max}\left(0.0106,1.2136-(-\Delta+h)^2\right)}$ &	8 & $0.0133$ & $0.6411$ \\
     & HS & $1.5561 - \sqrt{h}$ & 4 & $0.0594$ & $0.7931$ \\
    \Xhline{2\arrayrulewidth}
    \multirow{2}{4em}{$\sim 6000$}  & Best & $\sqrt{\mathrm{max}\left(0.0004,-h^2+(\Delta+0.9629)^2\right)}$ & 11 & $0.0102$ & $0.0065$ \\
    & HS & $1.5635 - \sqrt{h}$ & 4 & $0.0600$ & $0.7943$ \\
    \Xhline{2\arrayrulewidth}
    \end{tabular}
    \caption{The optimal equations for FS of the $XYZ$ model with the highest score (HS) and Best equations tabulated for several training sizes. The corresponding complexity, loss, and score are also presented. As the training size increases, the Best equation becomes closer to the actual FS. The HS and Best equation are not equivalent as evident from the complexity, loss and functional form.}
    \label{tab:xyz_training_study_results}
\end{table*}

{\bf FS of the \(KSEA\) model}. In this system, there is an additional parameter, $\{k_i\}$ along with $\{h_i, \gamma_i\}$, and hence, the complexity in equation increases. By varying system parameters, the NN entanglement of the \(KSEA\) model  for $\epsilon<10^{-4}$ is illustrated in Fig. \ref{fig:KSEA_9_plot}(a). The conspicuous hemisphere indicates one additional degree of freedom in the FS equation of the $XY$ spin chain in the form of parameter $k$.  Again by comparing $\mathcal{S}$ and $\mathcal{L}$ for a fixed complexity, we arrive at the best expression for FS (Figs. \ref{fig:KSEA_9_plot}(d)-(e)), 
\begin{equation}
    \gamma=\sqrt{1-0.99998 \min(1.0,h^2+k^2)},
    \label{Eq:FS_KSEA_SR}
\end{equation}
for $\epsilon=10^{-4}$, which matches very well with the known FS equation. Further, in the limit $k\to 0$, the Eq. (\ref{Eq:FS_KSEA_SR}) reduces to Eq. (\ref{eq:XY_FS_PySR}) obtained for the $XY$ model. Thus, this provides a guiding method to gain insights into the predicted equations by evaluating them at various limits. There are, however, some sharp differences in the $\epsilon$-study of the $KSEA$ model when compared to the $XY$ spin chain (see Figs. \ref{fig:KSEA_9_plot}(b)-(c)). In the latter, HS and Best equation are identical barring some differences in the values of constants while for  the \(H_{KSEA}\) with $\epsilon\ge5\cdot10^{-4}$, the two metrics lead to  two  distinct equations. Even for the case of the standard $\epsilon=10^{-4}$, the HS equation is 
\begin{equation}
    \gamma=1.0003\sqrt{-0.99946h^2-0.99946k^2+1},
\end{equation}
where, interestingly, it does not involve any discontinuous operators. In Fig. \ref{fig:KSEA_9_plot}(b), the Best equation at each $\epsilon$ has a smaller interquartile range, suggesting lower variations in equations in each round. The HS equations in each round vary quite significantly with respect to each other, thus producing a larger variation as visible in the violin plot. This primarily happens due to the increased equation complexity and a poor tradeoff between score and loss (see Fig. \ref{fig:KSEA_9_plot}(c)). This study establishes that the Best metric is more reliable.

\begin{figure}
    \includegraphics[width=\linewidth]{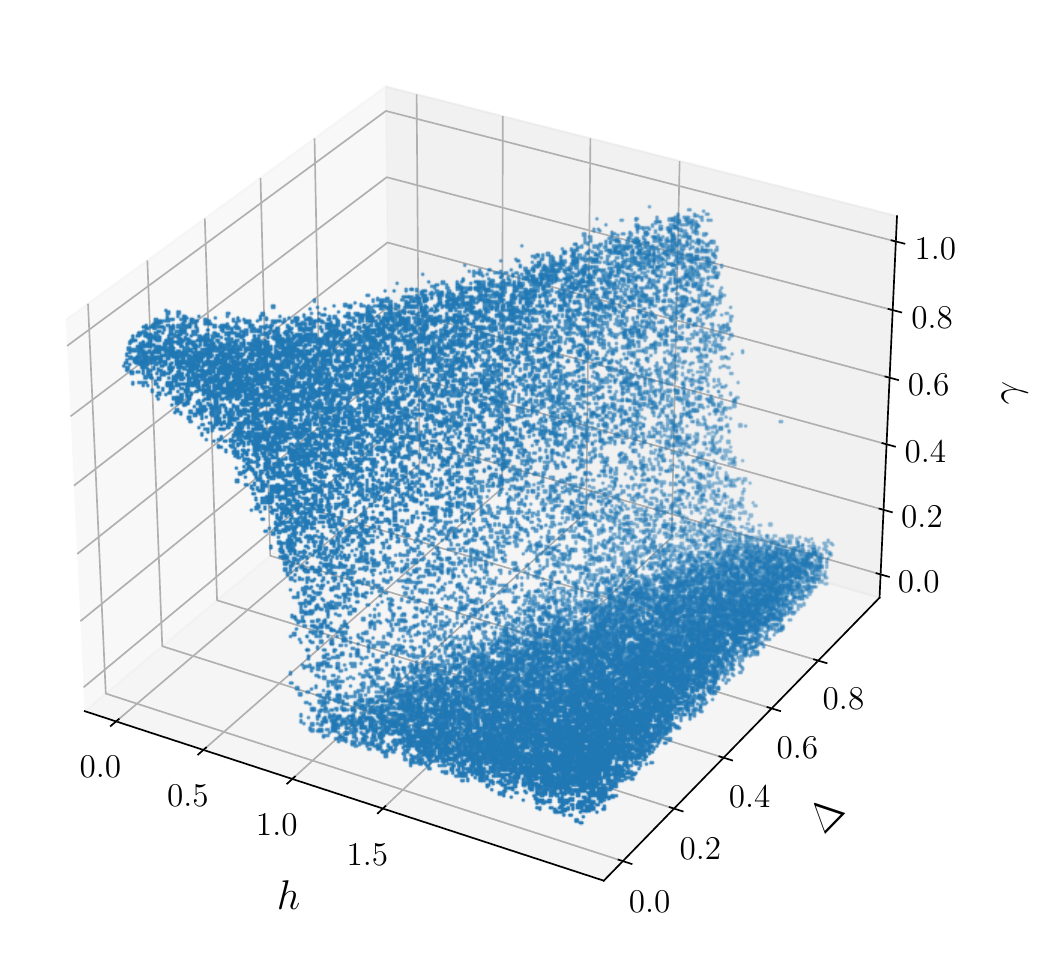}
    \caption{\textbf{Factorization surface of the \(XYZ\) model.} The surface with the Hamiltonian parameters when logarithmic negativity, $\mathcal{E} < \epsilon=5*10^{-2}$. $\mathcal{E}$ is generated via MPS ansatz for different parameters of the Hamiltonian, 
\(\Delta\) (\(x\)-axis), \(h\) (\(y\)-axis) and \(\gamma\) (\(z\)-axis). Other parameter of the system is \(N=60\). All axes are dimensionless.}
    \label{fig:xxz_plot}
\end{figure}

\textbf{FS of the non-integrable \(XYZ\) model}. Let us now move on to the quantum \(XYZ\) model in the presence of the external magnetic field, given in  Eq. (\ref{eq:XYZ_ham}). The model can be diagonalized only by numerical methods and hence has a different status than the ones discussed so far. Interestingly, however, the FS in terms of $h,~\gamma$ and $\Delta$ is known in the literature \cite{KURMANN1982235}. As there are density variations, i.e., the parameter space of entanglement $\mathcal{E}<\epsilon$ is not spread uniformly as shown in Fig. \ref{fig:xxz_plot}, we reach the predicted equations as a function of the size of the training data. We realize that  $4000$ points used for the \(KSEA\) model are too small here to obtain the right equation. This is due to the fact that \(KSEA\)'s hemispherical FS in the dataset has points with equal density distribution. In contrast,  since the density of the parameter space is smaller from $\gamma=0.2$ to $\gamma=0.6$ compared to the high $\gamma$ values for the \(XYZ\) dataset, the $\mathcal{L}$-metric tries to correct its course based on high-density regions and reduces the loss in these areas, compromising $\mathcal{L}$ in low-density areas. This leads to either wrong equations or the right equation which do not have the appropriate score-loss tradeoff and hence cannot be selected. We find that considering $\sim 6000$ training points, PySR predicts the equation 
\begin{equation}
    \gamma=\sqrt{\max\left(0.0004, - h^{2} + \left(\Delta + 0.9629\right)^{2}\right)}.
\end{equation}
\begin{table*}[]
    \centering
    \begin{tabular}{c @{\hspace{2em}} c@{\hspace{2em}} c@{\hspace{2em}} c}
    \Xhline{2\arrayrulewidth}
    \Xhline{2\arrayrulewidth}
    Equation & Complexity & Loss & Score \\ 
    \Xhline{2\arrayrulewidth}
    \Xhline{2\arrayrulewidth}
    $ 0.835$ & $1$ & $0.0357$ & $0.0$ \\\
    $\max\left(0.822, d\right)$ & $3$ & $0.0345$ & $0.0341$ \\\
    $0.995 - h^{2}$ & $4$ & $0.0207$ & $0.510$ \\\
    $(0.99510586 - h^2) +0.16635655$ & $6$ & $0.0119$ & $0.528$ \\\
    $\sqrt{0.995 - \min\left(0.995, h^{2}\right)}$ & $7$ & $1.45 \cdot 10^{-5}$ & $6.71$ \\\
    $\sqrt{\sqrt{0.99510586} - \min\left(0.995, h^{2}\right)}$ & $8$ & $5.81 \cdot 10^{-6}$ & $0.918$\\
    \Xhline{2\arrayrulewidth}
    \end{tabular}
    \caption{Predicted equations of the \(DUXY\) model for $d<\gamma$. We can observe that the equation with the highest score has a complexity of $7$ but a loss one order of magnitude higher than the``Best Equation" which is $\gamma=\sqrt{0.995-\min\left(0.995,h^2\right)}$}.
    \label{tab:duxy_d_less_than_gamma}
\end{table*}

This is in decent agreement with the analytic equation but the constant $0.96$ instead of unity can be optimized further either using template fitting or increasing training size as seen in  Table. \ref{tab:xyz_training_study_results}. One can observe that at each size of the training data, the HS metric forecasts a completely incorrect equation which is independent of $\Delta$. However, in case of the Best metric, we find that there is one more potential competing equation which has the sub-expression $(h-\Delta)^2$, dissimilar from the correct dependence $-h^2+(1+\Delta)^2$.  Additionally note that the system that does not have an analytical solution and can only be diagonalized through numerical procedures. Therefore, generating a large dataset is time-consuming and hence not possible practically. For the results of optimal equations at each $\mathcal{C}$ for $N_{\text{train}}\sim 6000$ (see Fig. \ref{fig:XYZ_9_plot_1}).

\subsection{Determining FS of the $DUXY$ model from PySR's algorithm}

{\it Ascertaining FS of $DUXY$ model with $d<\gamma$}. Having established that PySR can indeed predict FS of benchmark quantum spin chains where  expressions are known from analytical treatment, it is now time to test PySR on the NN  \(DUXY\) model for which the exact expression for FS is  unknown. For $d<\gamma$, the expression is independent of $d$ and identical to that of the $XY$ model. Two contrasting features are noticed in this case in Table \ref{tab:duxy_d_less_than_gamma}. which were not present in the $XY$ model -- (1) the equations at complexity $7$ and $8$ are similar and it can be attributed to the addition square root operation. (2) Corresponding to this complexity, loss suddenly drops off the order of $10^{-3}$, when compared to equation of complexity $6$ and at the same time, the score increases six times more, thereby illustrating the proper correlation between loss and score. 
In this case, the estimated expression turns out to be 
\begin{equation}
    \gamma=\sqrt{0.995-\mathrm{min}(0.995,h^2)},
    \label{eq:DUXY_FS_SR}
\end{equation}
which is also in good agreement with  Eq. (\ref{eq:XY_FS_PySR}). Notice, however, that since the dataset contains another set of parameters, the equations in the $XY$ and  in the $DUXY$ model with $d<\gamma$ are slightly different which only coincide with the known FS  provided the accuracy being two decimal places. 

\begin{figure}[H]
    \centering    \includegraphics[width=\linewidth]{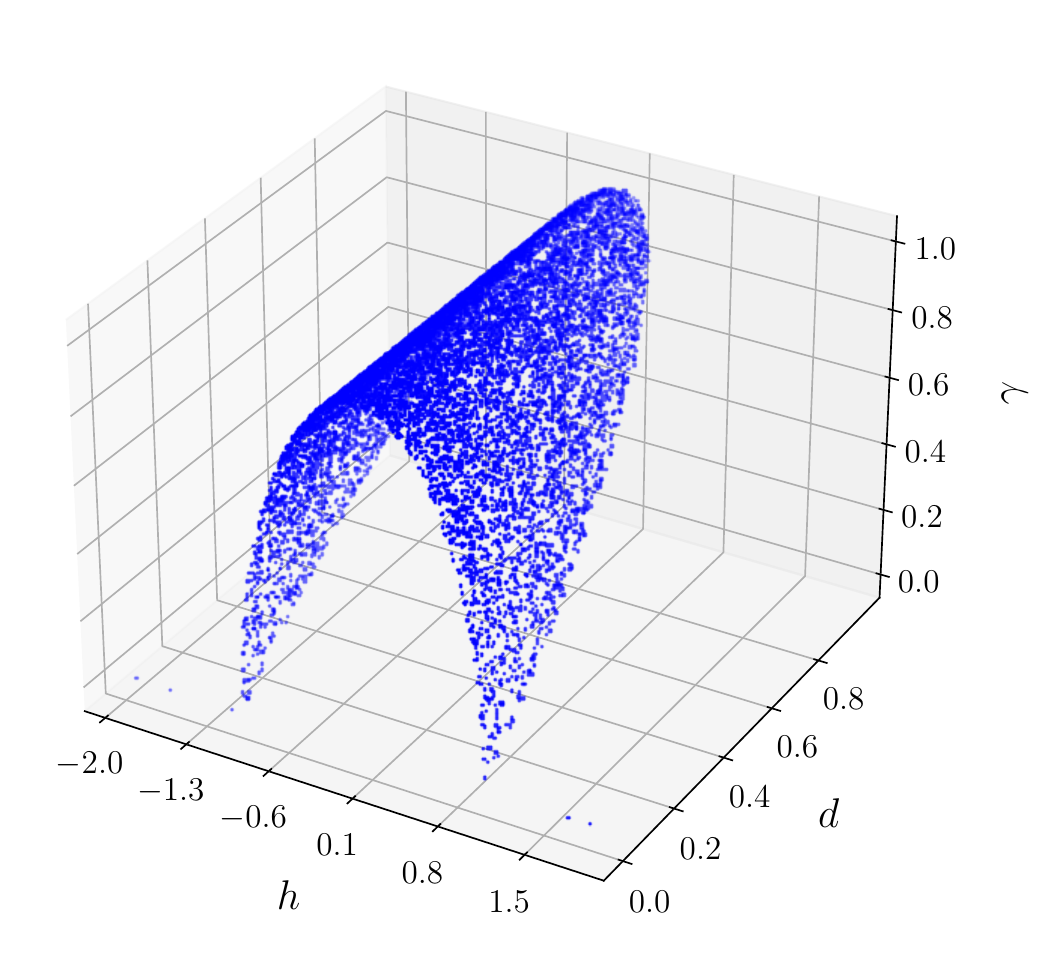}
    \caption{ \textbf{Factorization surface of the \(DUXY\) model.} We collect logarithmic negativity for $d<\gamma$, $\mathcal{E} < \epsilon=10^{-3}$ and plot the surface with the Hamiltonian parameters: \(h\) (\(x\)-axis), \(d\) (\(y\)-axis) and \(\gamma\) (\(z\)-axis). It is evident from   the shape of the factorization surface that its equation is independent of DM interaction, i.e., the strength of $d$. All  axes are dimensionless. }
    \label{fig:d<gamma_3dplot}
\end{figure}

\begin{figure*}  \includegraphics[width=\linewidth]{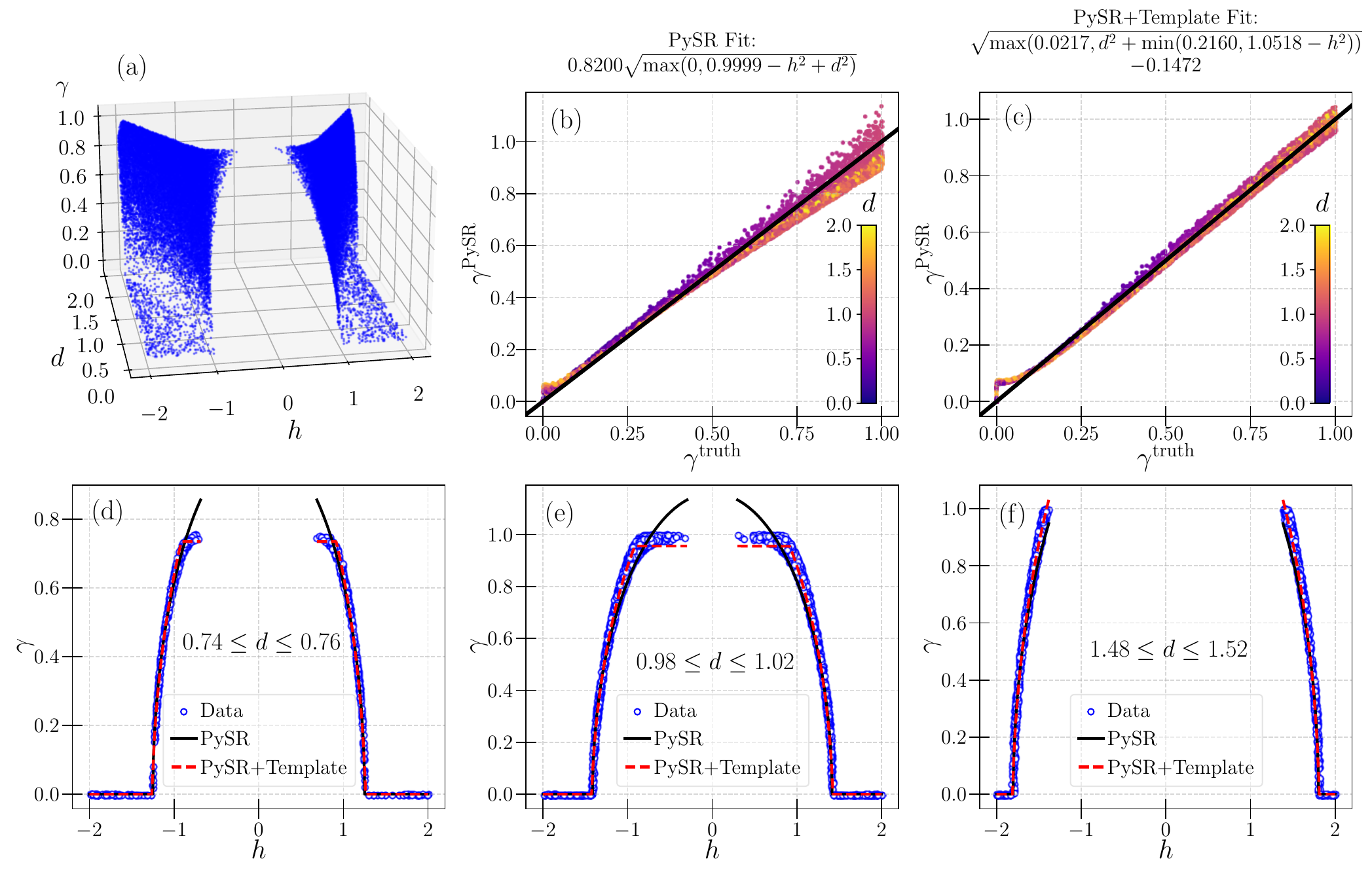}
    \caption{{Factorization surface of the \(DUXY\) model  with $d>\gamma$.} (a)  The surface of   $h$ (\(x\)-axis), $d$ (\(y\)-axis) and $\gamma$ (\(z\)-axis) in \(H_{DUXY}\) when \(\mathcal{E} < 5 * 10^{-4}\).  (b) PySR's preliminary prediction for $\gamma$ -- we can notice that the equation has a nice fit to the $y=x$ (black solid line). (c) After performing template fitting  by optimizing the constants along with PySR, the prediction improves overall  but deviates at low $\gamma$ values. It has a good fit in the middle and high values of $d$. (d)-(f) To illustrate the performance of PySR and template fitting in (b) and (c), $\gamma$ is plotted against $h$ for $0.74\le d\leq 0.76$, $0.98\le d\leq 1.02$ and $1.48\le d\leq 1.52$ respectively. Here, PySR's prediction overshoots the performance of the template fitting. We suspect that the fit has to be non-trivially non-linear, especially when  $\gamma \in [0.8,1.0]$ where PySR and template fit together perform very well. All axes are dimensionless.}
    \label{fig:DUXY_Model_initial_fits}
\end{figure*}

{\it Ascertaining FS of $DUXY$ model with $d>\gamma$}. In this domain, we encounter the difficulties of severe density variations. The relationship of $\gamma$ with $h$ and $d$ looks very non-linear as displayed in Fig. \ref{fig:DUXY_Model_initial_fits}(a). When we restrict the operator space to that of the $KSEA$ model, we find that the best equation is given by 
\begin{equation}
\gamma=0.82\sqrt{\max(0,0.9999-h^2+d^2)}.
    \label{eq:DUXY_FS_SR_wrong}
\end{equation}
This is what we call the ``PySR's Preliminary Prediction".
At first glance, this result might look tempting especially due to its simplicity and similarity to the FS of the \(XY\) model on taking $d\to 0$ as evident in Fig. \ref{fig:DUXY_Model_initial_fits}(c). One can notice that $\gamma^{\text{truth}}$ follows $\gamma^{\text{PySR}}$ along the $y=x$ line. At the same time, the scatter of the fit is higher for $\gamma^{\text{truth}}>0.75$. Further investigations (Fig. \ref{fig:DUXY_Model_initial_fits}(e)) disclose that for $0.95<d<1.05$, this equation overshoots the prediction and forms an ellipse, thereby indicating that this equation might not be a potential candidate for the \(DUXY\) model for the entire region with \(d>\gamma\). It was supposed to terminate when $\gamma>1$ and reduce to a non-linear discontinuous curve. This also informs that just plotting $\gamma^{\text{PySR}}-\gamma^{\text{truth}}$ alone is not sufficient to measure the correctness of an equation. One must also check its validity across various slices of the  parameter space of the model.

We also immediately realize that at the limit $d \rightarrow 0$,  Eq. (\ref{eq:DUXY_FS_SR_wrong}) does not reduce to Eq. (\ref{eq:XY_FS_PySR}). Note, however, when $\gamma>0$, $d \rightarrow 0$ is not a valid limit in the domain $d>\gamma$. This becomes evident in studying the entanglement pattern in the $(\gamma, h)$-plane for $d>\gamma$ (see Fig. \ref{fig:DUXY_Model_initial_fits}(a)). It can be observed that it predicts just the scaled-down equation of the \(XY\) model by a factor of $0.82$ and the other constants inside square root are valid up to the first decimal point. 
Now, we relax the operator space as well as nesting conditions and, thereafter, perform a template fitting to optimize the constants further. In this way, we reach to the final FS equation, given by
\begin{equation}   \gamma=\sqrt{\max(0.0217,d^2+\min(0.2160,1.0518-h^2)}-0.1472.   
\end{equation}
We call this equation as ``PySR+Template" Fit. In comparison to just PySR's prediction alone on restricted operator space, this equation has deviations from the $y=x$ line for $\gamma\in(0,0.1]$. This could, in principle, be an artifact of systematics like the density variations, numerical errors and statistics like number of training points. If one now investigates its performance along the $d=1$ slice in Fig. \ref{fig:DUXY_Model_initial_fits}(e), one can establish that the fit nicely cuts at $\gamma=1$. We also check the performance of both the methods along  $d=0.75, 1.5$ slices. In Fig. \ref{fig:DUXY_Model_initial_fits}(d), the PySR's preliminary prediction again surpasses the correct values at $|h|\sim 1$, whereas Template Fit matches perfectly. For the last slice in Fig. \ref{fig:DUXY_Model_initial_fits}(f), both the equations have decent fits. Overall, this suggests that using SR+Template Fitting as a potential tool, it may be possible to discover new interpretable empirical equations which can guide the existing theory.  

\subsection{Prediction of FS of the long-range \(XY\) spin model via SR }

For the long-range \(XY\) model, the ML techniques are utilized to obtain the FS equation for different coordination numbers, \(\mathcal{Z}\). The dataset, in this case, is generated with the help of exact diagonalization method. We illustrate that when the data is inserted to the PySR algorithm, it can predict factorization surface up to \(\mathcal{Z} =2\) with a high accuracy. 

For the case $\mathcal{Z}=2$,
the PySR  leads to an  equation, given by 
\begin{equation}
    \gamma^{\text{PySR}}=\sqrt{\max\left(0, \frac{h^{2}}{- 2.3998 R_{2} - 0.99299} + 0.99299\right)}. 
\end{equation}
where $R_2=1/2^{\alpha}$. On the other hand, the actual correct equation reads as 
\begin{equation}    \gamma^{\text{truth}}=\sqrt{\max\left(0, 1-\frac{h^{2}}{(1+R_2)^2}\right)}. 
\end{equation}
At first sight, it can appear that the SR's anticipated equation is  different from the actual one. However, a closer inspection of the sub-expression, $(1+R_2)^2=1+R_2^2+2R_2$, reveals that the prediction is correct to a linear order in $R_2$ in this sub-expression. Since $R_2$ is small, $R_2^2$ is very small and can be neglected. Thus, taking into account these factors, the two equations are similar in their functional form. 
If the  SR algorithm uses the dataset generated with \(\mathcal{Z}=3\) for arriving at the FS equation, we observe that the prediction does not work as good as we obtain for the NN and NNN models. It possibly suggests that for the LR model, certain modifications are required for arriving at the expression for the FS.

\section{Conclusion}
\label{sec:conclusion}

A closed-form expression that can  describe the available data is the aim of various physics endeavors. To that end, supervised machine learning (ML) techniques like symbolic regression can be employed. By using a symbolic regression software, called PySR, we arrived at the equation of factorization surface (FS), where the ground state of a particular Hamiltonian becomes fully  separable. In recent years, PySR has been effectively employed to obtain analytical solutions in numerous physics and mathematics domains. 

To confirm the PySR algorithm's capacity, we find the FS of the nearest-neighbor transverse $XY$ model with or without Kaplan-Shekhtman-Entin-Aharony (\(KSEA\)) interactions that matches the analytical expressions known in the literature.  In the $XY$ model with Dzyaloshinsky-Moriya (\(DM\)) interactions, there is a parameter regime in which FS is known, whereas in other circumstances, it is undetermined. In the former situation, PySR discovers the surface with a high accuracy, whereas in the latter case, symbolic regression produces a new equation that seems to match quite well.

Beyond the analytically solvable models, we address two spin chain models:  the nearest-neighbor \(XYZ\) model, which we solve using the density matrix renormalization group approach and the long-range \(XY\) model using Lanczos diagonalization method. While FS was suggested in the latter situation, it is known in the former case. We again found that the FS equation for the $XYZ$ model using the PySR approach coincides with an equation that was known. In contrast, the FS for the long-range  $XY$ model can only be predicted when the interactions involve nearest-neighbor and next nearest-neighbor sites. 
Our results not only validate the ML approach as a substitute strategy to derive analytical expression, characterizing features in many-body systems, but also highlight it as a potential candidate to unveil new physics that remains beyond the reach of analytical and numerical methods.



\section*{acknowledgements}

We acknowledge the support from Interdisciplinary Cyber Physical Systems (ICPS) program of the Department of Science and Technology (DST), India, Grant No.: DST/ICPS/QuST/Theme- 1/2019/23. We  acknowledge the use of \href{https://github.com/titaschanda/QIClib}{QIClib} -- a modern C++ library for general purpose quantum information processing and quantum computing (\url{https://titaschanda.github.io/QIClib}) and cluster computing facility at Harish-Chandra Research Institute. This research is supported in part by the “INFOSYS" scholarship for senior students.
\bibliography{references}

\appendix
\section{Analytic solution of $1D$ \(XY\) spin chain}
\label{app:xy_spin_model}
We  describe the diagonalization procedure for the one-dimensional \(XY\) Hamiltonian in presence of a transverse magnetic field \cite{lieb1961,barouch_pra_1970}. The Hamiltonian is given as
\begin{eqnarray}
    \nonumber H_{XY} &=& \sum_{i=1}^{N}\left [\frac{(1+\gamma)}{4}\sigma^x_i\sigma^x_{i+1}+\frac{(1-\gamma)}{4}\sigma^y_i\sigma^y_{i+1}\right ]  + \frac{h}{2}\sigma^z_i.
    \label{eq: XY_apprendix}
\end{eqnarray}
Let us define the spin ladder operators, $\sigma^+_i$ and $\sigma^-_i$, as
\begin{equation}
    \sigma^+_i = \frac{\sigma^x_i+i\sigma^y_i}{2}; \quad \sigma^-_i = \frac{\sigma^x_i-i\sigma^y_i}{2}\quad \forall i=1,2,\dots N.
    \label{s+-}
    \end{equation}
Thus the Hamiltonian  in terms of raising and lowering operators reduces to 
\begin{eqnarray}
   \nonumber H_{XY} &=& \sum\limits_{i=1}^{N}\Bigg[\frac{1}{2}(\sigma^+_i\sigma^-_{i+1}+\sigma^-_i\sigma^+_{i+1})\\&+&\frac{\gamma}{2}(\sigma^+_i\sigma^+_{i+1}+\sigma^-_i\sigma^-_{i+1}) 
     + h\sigma^+_i\sigma^-_i\Bigg].
    \label{XY}
\end{eqnarray}
We use the highly non-linear Jordan-Wigner transformation to represent $H_{XY}$ in terms of fermionic creation and annihilation operators, $c_i^\dagger$ and $c_i$ respectively, with    
\begin{equation}
    c_i = e^{-i\pi\sum\limits_{k=1}^{i-1}\sigma^+_k\sigma^-_{k}}\sigma^-_i; c^\dagger_i = \sigma^+_i e^{i\pi\sum\limits_{k=1}^{i-1}\sigma^+_k\sigma^-_{k}}.
\end{equation}
In the thermodynamic limit (i.e., $N \rightarrow \infty $), the boundary term would have infinitesimal contribution and hence, ignoring the boundary term, we obtain
\begin{eqnarray}
    \nonumber H_{XY} &=& \sum\limits_{i=1}^{N}\Bigg[\frac{1}{2}(c^\dagger_i c_{i+1}+c_i c^\dagger_{i+1})\\ &+&\frac{\gamma}{2}(c^\dagger_ic^\dagger_{i+1}+c_i c_{i+1})
     + h c^\dagger_i c_i\Bigg].
    \label{XYJW}
\end{eqnarray}
We use the Fourier transform of the fermionic operators (because of the translation invariance), given by
\begin{equation}
    a_p^\dagger = \frac{1}{\sqrt{N}}\sum\limits_{j=1}^{N}e^{-i j\phi_p} c^\dagger_j; a_p = \frac{1}{\sqrt{N}}\sum\limits_{j=1}^{N}e^{i j\phi_p} c_j, 
    \label{XYFT}
\end{equation}
with $\phi_p=\dfrac{2\pi p}{N}\quad\forall p = [-\frac{N}{2}, \frac{N}{2}]$. Combining $\pm p$, $H_{XY}$ is decoupled in the basis, $\{|0\rangle, a^\dagger_p a^\dagger_{-p}|0\rangle, a^\dagger_p|0\rangle,a^\dagger_{-p}|0\rangle\}$, i.e.,
\begin{equation}
    H_{XY} = \sum\limits_{p=1}^{N/2}I\otimes I\otimes\cdots\otimes H^p_{XY}\otimes\dots\otimes I,
\end{equation}
with $I$ being a $4\times4$ identity matrix and
\begin{equation}
\begin{aligned}
    H^p_{XY}&=\left[\begin{array}{cccc}
h & -i \gamma s_p & 0 & 0 \\
i \gamma s_p  & 2c_p-h & 0 & 0 \\
0 & 0 & c_p & 0 \\
0 & 0 & 0 & c_p\\
\end{array}\right], 
\end{aligned}
\end{equation}
where $c_p = \cos \phi_p$ and $s_p = \sin \phi_p$.
Any two-site density matrix can be represented as
\begin{equation}
    \rho_{i,j}=\frac{1}{4} \left(
    \mathbb{I}_4 + \vec{r} \cdot \vec{\sigma}  \otimes \mathrm{I} + \mathbb{I} \otimes \vec{s} \cdot \vec{\sigma} + \sum_{k,l = x,y,z} C_{kl} \sigma_{k} \otimes \sigma_{l} \right),
\end{equation}
where \(C_{kl}=\tr(\sigma_{k} \otimes \sigma_{l} \rho_{i,j}),\hspace{0.05in} k,l =x,y,z\) and \(\vec{r}\), \(\vec{s}\) denote the magnetization vector. 
The two-site density matrix in this case can be represented as
\begin{equation}
    \rho_{i,j}=\frac{1}{4} \left (\mathbb{I}_4 + m^z({\sigma^z_i}+{\sigma^z_{j}}) + \sum_{k=x, y,z} C^{k}_{i,j} \sigma^{k}_i \otimes \sigma^{k}_{j}\right).
\end{equation}
When additional \(KSEA\) and \(DM\) interactions are included in the  \(XY\) model, the above analytical treatment can also be applied and the form of  \(\rho_{i,j}\) remains same except there are two more nonvanishing correlators, \(C_{i,j}^{xy}\) and  \(C_{i,j}^{yx}\). \\
\begin{figure*}
    \centering
    \includegraphics[width=\textwidth]{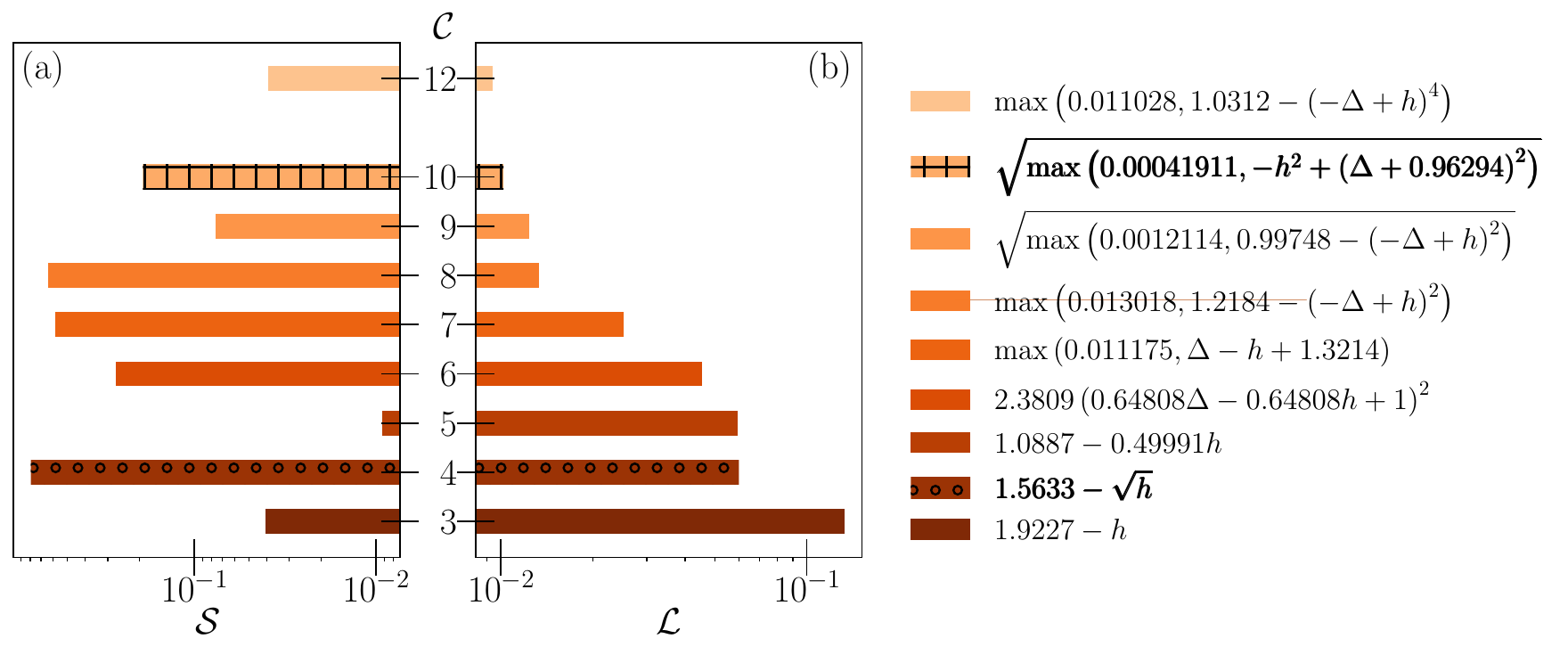}
    \caption{Symbolic regression results for the FS of the  \(XYZ\) Model with $N_{\text{train}}\sim 6000$. The equation in bold $1.5633-\sqrt{h}$ refers to the HS equation, while the other bold equation is the Best equation as discussed in the main text.}
    \label{fig:XYZ_9_plot_1}
\end{figure*}

\section{Factorization Surface}
\label{app:fact_surface_calc}

 The  \(KSEA\) model with periodic boundary condition (PBC) is described as \cite{kaplan_jpb_1983}
\begin{eqnarray}
    \nonumber H_{KSEA}&=&\sum_{r}\frac{(1+\gamma)}{4}\sigma_r^x\sigma_{r+1}^x+\frac{(1-\gamma)}{4}\sigma_r^y\sigma_{r+1}^y\\&&+\frac{k}{4}(\sigma_r^x\sigma_{r+1}^y+\sigma_r^y\sigma_{r+1}^x)+\frac{h}{2}\sigma_r^z,
\end{eqnarray}
where \(k\) is the coupling constant of the \(KSEA\) interactions. 
By tuning the parameters, we can obtain the ground state of this model analytically by using the procedure described in the previous section. It can be shown to be doubly degenerate and does not possess any bipartite and multipartite entanglement. 
This leads to the equation of factorization surface, which depends on the Hamiltonian parameters as
\begin{equation}
    \gamma^2=1-h^2-k^2.
\end{equation}
In the \(XY\) model with \(k=0\), the FS reduces to $h^2+\gamma^2=1.$


\end{document}